
\input harvmac

\def\dag{\dagger}
\def\VEV{VEV}
\noblackbox
\baselineskip=12pt

\Title{\vbox{\baselineskip12pt{\hbox{CTP-TAMU-26/93}}%
{\hbox{hep-ph/9305285}}}}
{\vbox{\centerline{Electroweak Z-string and $\tan\beta$\footnote{$^\dagger$}{%
Talk presented at SUSY '93, March 29-April 1st, Northeastern University,
Boston, MA 02115, USA.}}}}


\vskip.4truein

\centerline{HoSeong ~La\footnote{$^*$}{%
e-mail address: hsla@phys.tamu.edu, hsla@tamphys.bitnet}   }

\bigskip\centerline{Center for Theoretical Physics}
\centerline{Texas A\&M University}
\centerline{College Station, TX 77843-4242, USA}
\vskip 0.3in

\midinsert
\narrower
In the standard model with two Higgs doublets
it is shown that the existence of the electroweak Z-string in general
requires the same characteristic length for the two Higgs fields as well as
a specific ratio of the two Higgs vacuum expectation values,
i.e. $\tan\beta$.
This ratio can be determined in terms of the couplings in the Higgs potential.
Some remarks on the supersymmetric case are given.
\endinsert

\baselineskip=12pt

 \noblackbox
\nopagenumbers

\def\hat{\widehat}
\def\tilde{\widetilde}

\def\la{\lambda}
\def\gt{{\tilde g}}
\def\half{{\textstyle{1\over 2}}}

\def\e{{\rm e}}
\def\pa{\partial}
\def\mbox#1#2{\vcenter{\hrule \hbox{\vrule height#2in
		\kern#1in \vrule} \hrule}}  
\def\lfr#1#2{{\textstyle{#1\over #2}}}
\def\eps{\epsilon}

\def\rbr{\right)}
\font\cmss=cmss10 \font\cmsss=cmss10 scaled 833
\def\IZ{\relax\ifmmode\mathchoice
{\hbox{\cmss Z\kern-.4em Z}}{\hbox{\cmss Z\kern-.4em Z}}
{\lower.9pt\hbox{\cmsss Z\kern-.4em Z}}
{\lower1.2pt\hbox{\cmsss Z\kern-.4em Z}}\else{\cmss Z\kern-.4em Z}\fi}
\def\IR{\relax\ifmmode\mathchoice
{\hbox{\cmss I\kern-.5em I}}{\hbox{\cmss R\kern-.5em R}}
{\lower.9pt\hbox{\cmsss I\kern-.5em I}}
{\lower1.2pt\hbox{\cmsss R\kern-.5em R}}\else{\cmss R\kern-.5em R}\fi}
\def\tr{{\rm tr}}

\def\cos{{\rm cos}}
\def\sin{{\rm sin}}

\def\CL{{\cal L}}


\newsec{Introduction}

The idea of the spontaneous symmetry breaking is one of the most important
building blocks of the electroweak theory, which is one of the greatest
achievements in theoretical particle physics.
Although there is no doubt about the validity of the electroweak theory,
the final experimental proof however is not yet completed because
the wanted Higgs particle is still at large. And also there is some possibility
that the detail of such spontaneous symmetry breaking may be a bit different
from the minimal content originally proposed.
There is growing anticipation, although indirect, the electroweak symmetry
breaking may be induced by two Higgs doublets, rather than one.

For example,
the recent measurements of the gauge couplings\ref\PDG{Particle Data Group,
Phys. Rev. {\bf D45} (1992).}\ have led us to anticipation
that one of the minimal supersymmetric Grand Unified Theories
(GUTs)\ref\rSGUT{S.
Dimopoulos and H. Georgi, Nucl. Phys. {\bf B193} (1981) 150; N. Sakai, Zeit. f.
Phys. {\bf C11} (1981) 153.}\
or supergravity GUTs\ref\rSGGT{A.H. Chamseddine, R. Arnowitt and P. Nath, Phys.
Rev. Lett. {\bf 49} (1982) 970; L.E. Ibanez, Phys. Lett. {\bf 118B} (1982) 73;
J. Ellis, D.V. Nanopoulos and Tamvakis, Phys. Lett. {\bf 121B} (1983) 123;
K. Inoue, A. Kakuto, H. Komatsu and S. Takeshita, Prog. Theo. Phys. {\bf 68}
(1982) 927; L. Alvarez-Gaum\'e, J. Polchinski and M.B. Wise, Nucl. Phys. {\bf
B221} (1983) 495; J. Ellis, J.S. Hagelin, D.V. Nanopoulos and Tamvakis, Phys.
Lett. {\bf 125B} (1983) 275; L. Iba\~nez and C. Lopez, Nucl. Phys. {\bf B233}
(1984) 545; L.E. Iba\~nez, C. Lopez and C. Mu\~nos, Nucl. Phys. {\bf B250}
(1985) 218.}\ref\revSGUT{For reviews see P. Nath, R. Arnowitt and A.H.
Chamseddine, ``{\it Applied N=1 Supergravity}," (World Sci., Singapore,
1984); H.P. Nilles, Phys. Rep. {\bf 110} (1984) 1; H. Haber and G. Kane, Phys.
Rep. {\bf 117} (1985) 75.}\
with the supersymmetry scale of order 1 TeV or below may be a
phenomenologically plausible unified theory of strong and electroweak
interactions\ref\rUNCoup{P. Langacker, in {\it Proc. PASCOS 90 Symposium} ed.
by P. Nath and S. Reucroft (World Sci., Singapore, 1990); P. Langacker and M.
Luo, Phys. Rev. {\bf D44} (1991) 817; J. Ellis, S. Kelley and D.V. Nanopoulos,
Phys. Lett. {\bf 249B} (1990) 441; {\bf 260B} (1991) 131; U. Amaldi, W. de Boer
and H. F\"ursteanu, Phys. Lett. {\bf 260B} (1991) 447.}.
These supersymmetric GUTs in general require at least two Higgs
multiplets for the electroweak symmetry breaking\revSGUT%
\ref\rtwoH{N.G. Deshpande and E. Ma, Phys. Rev. {\bf D18} (1978) 2574;
 J.F. Donoghue and L.-F. Li, Phys. Rev.
{\bf D19} (1979) 945; H.E. Haber, G.L.Kane and T. Sterling, Nucl. Phys. {\bf
B161} (1979) 493; E. Golowich and T.C. Yang, Phys. Lett. {\bf 80B} (1979) 245.
}\ref\GerH{H. Georgi, Hadronic J. {\bf 1} (1978)
155.}\ref\revHig{For a review see J.F. Gunion, H.E. Haber, G. Kane and S.
Dawson, ``{\it The Higgs Hunter's Guide}," (Addison-Wesley, 1990).}.
The recent studies of the baryogenesis also favors going beyond the one Higgs
doublet case.

In exchange of these positive points, having one more Higgs doublet will
introduce further  complication to the theory. Needless to say, first,
we have to deal with more observable massive scalar particles, despite having
difficulties to find single Higgs particle.
Theoretically, it also introduces more free parameters.
To spontaneously break the $SU(2)\times U(1)_Y$ symmetry
down to the $U(1)_{em}$ each Higgs
gets its own vacuum expectation value (\VEV), say $v_1, v_2$.
These \VEV s are phenomenologically important
but unfortunately they are not determined theoretically except in some no-scale
models\ref\rNano{A.B. Lahanas and D.V. Nanopoulos, Phys. Rep. {\bf 145} (1987)
1.}. The geometric sum $v^2/2=v_1^2+v_2^2$ can be determined in terms of the
mass of the gauge boson, where $v$ denotes the electroweak symmetry breaking
scale. This however leaves the ratio of the two \VEV s,
$\tan\beta\equiv v_2/v_1$,  still undetermined.

If the two-doublet model would turn out to explain the electroweak symmetry
breaking, eventually future experiments will determine $\tan\beta$. That will
however still leave us a question why this case should be different from
others.
We always wonder if nature selects out some particular property among many
possible choices. Then curiosity drives us to look for some explanation.
This is fairly a common situation. Pursuing the answer to such a question,
we are often led to a new phenomenon in physics.
Thus it is very important to look for any argument to constrain the
ratio rather theoretically, if possible.

With such a motivation in mind, in this talk I shall attempt to find any
relation to constrain  $\tan\beta$ in two-Higgs systems.
The result is indeed positive
and we find that there is a simple formula to express $\tan\beta$ in terms of
the couplings of the Higgs potential, so far as nature admits certain
vacuum defects during the electroweak phase transition.
The details are presented in
\ref\Mymhig{H.S. La, Texas A\&M preprint, CTP-TAMU-73/92 (hep-ph/9211215)
(1992).}%
\ref\Myzstring{H.S. La, Texas A\&M  preprint, CTP-TAMU-84/92 (hep-ph/9212286)
(1992).}%
\ref\Mydhiggs{H.S. La, Texas A\&M preprint, CTP-TAMU-1/93 (hep-ph/9302220)
(1993).}.

It was pointed out that the standard model with one Higgs doublet
admits string-like solution%
\ref\Namb{Y. Nambu, Nucl. Phys. {\bf B130} (1977) 505.}%
\ref\HuTi{K. Huang and R. Tipton, Phys. Rev. {\bf D23} (1981) 3050.}%
\ref\Vach{T. Vachaspati, Phys. Rev. Lett. {\bf 68} (1992) 1977;
T.~Vachaspati and M.~Barriola, Phys. Rev. Lett. {\bf 69} (1992) 1867.}.
In this case, there is no obvious topological configuration because the
hypercharge $U(1)_Y$ is not spontaneously broken, but rather
$SU(2)\times U(1)_Y$ is broken to $U(1)_{em}$.
In ref.\Namb, however, a solution of
a pair of $SU(2)$ magnetic monopoles connected by a string is derived.
(A similar solution can also be obtained in our case\Mydhiggs.)
It was also pointed out that in the standard model case these monopoles can
promote the stability of this system.
One can also derive string-like solutions in the
standard model with one Higgs doublet without
attaching to monopoles\Vach. These are unstable solutions except for
$\sin\theta_W=1$, where $\theta_W$ is the Weinberg angle.

In this talk we shall extend the results to the cases with more Higgs scalars.
As soon as we introduce more scalar fields, the potential to induce spontaneous
symmetry breaking becomes more complicated. In particular there are more
couplings including the interactions between different scalar fields.
The structure of the phase transition itself may become more involved.
For example, in a simpler case without gauge couplings and
only two scalar couplings there are already three different critical points%
\ref\Amit{D.J. Amit, ``{\it Field Theory, the Renormalization Group,
and Critical Phenomena}," (2nd ed.) (World Scientific, 1984).}.
If the two VEVs are very much different, each scalar can get its VEV one by
one.
In the two-doublet model case, if so,
we can naively anticipate that the phase transition
would occur in two steps.
Thus perhaps a full investigation of the structure of the
fixed points of two-Higgs potential may be necessary.


\newsec{Two-Higgs-Doublet Standard Model}

One of the most mysterious parts of the electroweak theory lies in the Higgs
sector. Higgs was introduced to achieve the electroweak symmetry breaking
without spoiling the consistency of the theory.
In this section we shall investigate the structure of a string-like defect
(so-called ``Z-string"), which can be formed during the electroweak phase
transition, in the two-Higgs-doublet standard model.

Note that the topology of the local $SU(2)\times U(1)/U(1)$ is the same as that
of $SU(2)$, so $\pi_1(SU(2)\times U(1)/U(1)) = 0$. As is well known, thus
the string-like solitonic solutions we get would not be topological.\foot{
$U(1)$ case is topological.}
However it is known that there are
nontopological solitons in the standard model\Namb\HuTi\Vach.

We shall use the CP invariant two-doublet Higgs potential that
induces $SU(2)\times U(1)_Y\to U(1)_{em}$ symmetry breaking\GerH\revHig :
\eqn\edhigpot{\eqalign{
V(\phi_1,\phi_2)=&\lfr{1}{2}\la_1\left(|\phi_1|^2-v_1^2\right)^2+
\lfr{1}{2}\la_2\left(|\phi_2|^2-v_2^2\right)^2+
\lfr{1}{2}\la_3\left(|\phi_1|^2+|\phi_2|^2-v_1^2-v_2^2\right)^2 \cr
&+\la_4\left(|\phi_1|^2 |\phi_2|^2-|\phi_1^\dag\phi_2|^2\right)
+\la_5\left|\phi_1^\dag\phi_2 -v_1 v_2\right|^2,\cr}}
where $\phi_1,\ \phi_2$ are $SU(2)$ doublets.
In this talk for argument's sake I shall take a simpler case
that $\la_i> 0$ for $i=1,2,3$ and assume $\la_4=\la_5=0$. (Later I shall quote
the result of the $\la\neq 0$ case.)
This potential shows $\phi_1, \phi_2\leftrightarrow -\phi_1, -\phi_2$
discrete symmetry, which is
necessary to suppress the flavor changing neutral current.
Then we shall find that this system reveals a rather interesting result,
which cannot be obtained otherwise.

Note that if $\la_4=0=\la_5$, the Higgs potential has a global $U(2)\times
U(2)$ symmetry. The symmetry breaking will lead to global
$U(1)\times U(1)$ unbroken
so that there will be two Goldstone bosons left over. If only $\la_5$ is
vanishing, due to $|\phi_1^\dagger\phi_2|^2$ in the $\la_4$-term
the global symmetry now becomes
$U(2)\times U(1) \times U(1)$. There still is one Goldstone boson after
symmetry breaking to global $U(1)\times U(1)$.

In reality there however are good reasons to keep $\la_5\neq 0$.
First, we don't want Goldstone bosons which may not be welcomed
phenomenologically. Secondly, though not directly related,
in the supersymmetric models $\la_5$ is related to the supersymmetry breaking
parameter so that $\la_5\neq 0$ to have the supersymmetry broken.
Note that for $\la_5\neq 0$ there is a global $U(2)$ symmetry, which is locally
isomorphic to $SU(2)\times U(1)$.
Nevertheless, in this talk just for convenience we set $\la_4=\la_5=0$.
Later we shall present the result of the general case too.
Let us consider the bosonic sector of the standard model
described by the Lagrangian density
\eqn\edlagr{
\CL=-\half\tr G_{\mu\nu}G^{\mu\nu}-{\textstyle{1\over 4}}F^{\mu\nu}F_{\mu\nu}
+|D_\mu\phi_1|^2+|D_\mu\phi_2|^2
	-V(\phi_1, \phi_2),}
where $F_{\mu\nu}=\pa_\mu B_\nu-\pa_\nu B_\mu$,
$G_{\mu\nu}^a=\pa_\mu W_\nu^a-\pa_\nu W_\mu^a + g\eps^{abc} W_\mu^b W_\nu^c$,
and $D_\mu=\pa_\mu-ig'{Y\over 2}B_\mu-ig{\tau^a\over 2} W_\mu^a$.
Both Higgs' have hypercharge $Y=1$.

Then the equations of motion for the scalar fields are
\eqna\edomi
$$\eqalignno{
0=&D^\mu D_\mu\phi_1 +\la_1\left(|\phi_1|^2-v_1^2\right)\phi_1
	+\la_3\left(|\phi_1|^2+|\phi_2|^2-v_1^2-v_2^2\right)\phi_1,&\edomi a\cr
0=&D^\mu D_\mu\phi_2 +\la_2\left(|\phi_2|^2-v_2^2\right)\phi_2
	+\la_3\left(|\phi_1|^2+|\phi_2|^2-v_1^2-v_2^2\right)\phi_2,&\edomi b\cr}
$$
and for the gauge fields we have
\eqna\edomic
$$\eqalignno{
&-\pa^\mu F_{\mu\nu}=j_\nu\equiv j_{1\nu} + j_{2\nu}, &\edomic a\cr
&\quad j_{i\nu} \equiv\lfr{1}{2}ig' \left[\phi_i^\dag\pa_\nu\phi_i
-(\pa_\nu \phi_i)^\dag\phi_i\right]
+\lfr{1}{2}g'^2 B_\nu |\phi_i|^2+\lfr{1}{2}g'g W_\nu^a\phi_i^\dag\tau^a\phi_i,
\ \ i=1,2,&\cr
&-\pa^\mu G^a_{\mu\nu} -g\eps^{abc}W^{b\mu}G^c_{\mu\nu}
=J^a_\nu\equiv J^a_{1\nu} + J^a_{2\nu}, &\edomic b\cr
&\quad J^a_{i\nu} \equiv\lfr{1}{2} ig \left[\phi_i^\dag\tau^a\pa_\nu\phi_i
-(\pa_\nu \phi_i)^\dag\tau^a\phi_i\right]
+\lfr{1}{2}gg' B_\nu \phi_i^\dag\tau^a\phi_i
+\lfr{1}{2}g^2W_\nu^b\phi_i^\dag\tau^a\tau^b\phi_i,
\ \ i=1,2, \cr}
$$
For time-independent solutions we choose $B_0=0=W_0^a$ gauge and impose the
cylindrical symmetry around the string, then the system
effectively reduces to a two-dimensional one. In this case the string solutions
in the $(1+3)$-dimensional spacetime correspond to
the vortex solutions in $\IR^2$.
When Higgs gets \VEV, the false vacuum region forms vacuum defects.
As usual,  we redefine the neutral gauge fields as
\eqn\azfld{
A_\mu=\cos\theta_W B_\mu+\sin\theta_W W_\mu^3,\quad
Z_\mu=\sin\theta_W B_\mu-\cos\theta_W W_\mu^3,}
where $\theta_W$ is the Weinberg angle defined by $\tan\theta_W=g'/g$.
We shall also use $\gt\equiv \half\sqrt{g^2+g'^2}$ for convenience.

For vortex solutions
it is convenient to represent them in the polar coordinates
$(r, \theta)$\ref\NieOl{H.B. Nielsen and P. Olesen,
Nucl. Phys. {\bf B61} (1973) 45;
E.B. Bogomol'nyi, Sov. J.
Nucl. Phys. {\bf 24} (1976) 449; H.J. de Vega and F.A. Schaposnik, Phys. Rev.
{\bf D14} (1976) 1100.}\ such as
\eqn\edans{\phi_1=\pmatrix{0\cr\e^{im\theta} f_1(r)\cr},
\quad \  \phi_2=\pmatrix{0\cr \e^{in\theta} f_2(r)\cr},\quad \
	{\vec Z}={\hat e_\theta} {1\over r}Z(r),}
where $m,n$ are integers identifying each ``winding'' sector.
Here we are mainly interested in the case of $W_\mu^1=0=W_\mu^2$,
but we expect there are other solutions
similar to the case of ref.\Vach.

Then Eqs.\edomic{a,b}\ become
\eqna\edmiia
$$\eqalignno{
0=&-{1\over r}\pa_r(r\pa_r B_\theta)+{1\over r^2}B_\theta &\cr
&-{g'\over r}\left[\left(m-\half(g'B_\theta-g W_\theta^3)\rbr f_1^2
+ (n-\half(g'B_\theta-g W_\theta^3) )f_2^2\right] , &\edmiia a\cr
0=&-{1\over r}\pa_r(r\pa_r W^3_\theta)+{1\over r^2}W^3_\theta &\cr
&+{g\over r}\left[\left(m-\half(g'B_\theta-g W_\theta^3)\rbr f_1^2
+ (n-\half(g'B_\theta-g W_\theta^3) )f_2^2\right] . &\edmiia b\cr}
$$
As we can easily see, $A_\mu$ satisfies a trivial equation so that we can
set $A_\mu=0$. Thus from the rest of the equations of motion we obtain
\eqna\edomiia
$$\eqalignno{
0&=-{1\over r}\pa_r(r\pa_r f_1) +{1\over r^2}f_1(m-\gt Z)^2
+ (\la_1 +\la_3) (f_1^2 - v_1^2)f_1
+\la_3 (f_2^2- v_2^2)f_1, \quad\quad&\edomiia a\cr
0&=-{1\over r}\pa_r(r\pa_r f_2) +{1\over r^2}f_2(n-\gt Z)^2
+ (\la_2 +\la_3) (f_2^2- v_2^2)f_2 +\la_3 (f_1^2- v_1^2)f_2, &\edomiia b\cr
0&=-\pa_r^2 Z+{1\over r}Z
-2\gt\left[(m-\gt Z) f_1^2 + (n-\gt Z)f_2^2\right]. &\edomiia c\cr}
$$

To become  desired finite-energy defects located at $r=0$
the solutions we are looking for should satisfy the
following boundary conditions:
\eqn\ebc{\eqalign{
f_1(0)=0,\ \ f_2(0)=0,\ \ &Z(0)=0,\cr
f_1\to v_1,\ f_2\to v_2, \ &Z\to {\rm const.}\ \ {\rm as}\ \
r\to\infty.\cr}}
The constant for the asymptotic value of $Z$ will be determined properly later.

We shall look for asymptotic solutions.
Imposing the boundary conditions at large $r$,
Eqs.\edomiia{a,b}\
become consistent only if $m=n$ and that it fixes the asymptotic value
$ Z\to {n/ \gt}$ as $r\to\infty$.
This implies that there is no vortex solution of different ``winding''
numbers for different Higgs fields.
With this condition of winding numbers we can solve
Eq.\edomiia{c}\ for large $r$ to obtain\NieOl\
\eqn\dsolA{
Z\to {n\over \gt}-n\sqrt{{\pi v\over 2\gt}}{\sqrt r} \e^{-r/\la}+\cdots,}
where $\la=1/\gt v$ is the characteristic length of the gauge field.
The coefficient of the second term is determined from the exact solution of
Eq.\edomiia{c}\ with $f_i^2=v_i^2$.
Note that the characteristic length defines the region over which the field
becomes significantly different from the value at the location of the defect.

The asymptotic solutions for $\phi_1$ and $\phi_2$  can be found as follows:
For simplicity we consider $n=1$ case, but the result does not
really depend on $n$. Besides, since these are nontopological, it is not really
necessary to consider other $n$ sector.
Asymptotically we look for solutions of the form
\eqn\edI{
f_1 -v_1\sim c_1v_1\e^{-r/\xi_1},\ \ f_2-v_2 \sim c_2v_2\e^{-r/\xi_2},}
where $\la_1$ and $\la_2$ are the characteristic lengths of
$\phi_1$ and $\phi_2$ respectively and
the constant coefficients $c_1$ and $c_2$  are in principle calculable, thus
they are not free parameters. Most likely these will appear as integration
constants, which would be fixed by the boundary conditions.
Let $c\equiv c_1/c_2$, then in the leading order we obtain
\eqna\eIIa
$$\eqalignno{
v_1\e^{-r/\xi_1}
\left[-{1\over \xi_1^2}+2(\la_1+\la_3)v_1^2
\right] +2{1\over c}\la_3 v_1 v_2^2
\e^{-r/\xi_2}+\cdots &=0,\quad \quad \ \  &\eIIa a\cr
v_2\e^{-r/\xi_2}
\left[-{1\over \xi_2^2}+2(\la_2+\la_3)v_2^2
\right] +2c\la_3 v_1^2 v_2
\e^{-r/\xi_1}+\cdots &=0,\ \ \ \ &\eIIa b\cr}
$$
where the ellipses include terms which vanish more rapidly for large $r$.

Recall that $\la_3>0$.
Thus to have any vortex solution we are forced to identify the two
characteristic lengths of the scalar fields such that
$$\xi\equiv\xi_1=\xi_2.$$

\def\tc{c}
Then the condition for
$\xi_1=\xi_2$ reads as
\eqn\enori{\tan\beta=\sqrt{{\la_1+\la_3-\tc \la_3\over \la_2+\la_3 - {1\over
\tc}\la_3}}.}
For $c=1$ we obtain
\eqn\eresult{\tan\beta=\sqrt{{\la_1\over\la_2}}.}

Now we shall show explicitly why Eq.\enori\ is not a formula to determine $c$
for $\tan\beta$ as an input parameter.
The  main rationale is that such a  constant is
not usually determined by solving nonlinear differential equations
asymptotically. Thus one should use it as an input parameter. Of course it
will be determined by solving equations exactly, however we are not yet able to
solve the equations in question exactly.
As we shall see, this constant will be ill-defined for Eq.\edI\ to define the
characteristic length unambiguously, if we try to determine
using Eq.\enori.

Suppose $\tc$ is not an input parameter but were to be determined by this
equation for any $\tan\beta$ to nullify our claim that this equation relates
$\tan\beta$ and other couplings, then one should be able to determine $\tc$ for
given $\tan\beta$.
In this case eq.\enori\ becomes a quadratic equation for $\tc$ as
\eqn\enorii{\la_3 \tc^2 + \left(\tan^2\beta(\la_2+\la_3) -
(\la_1+\la_3)\right)\tc -\la_3\tan^2\beta=0}
and one can obtain
\eqn\enoiii{\tc_\pm\!=\!{1\over 2\la_3}\left[-\tan^2\!\!\beta(\la_2\!+\!\la_3
\!)\!+\! (\la_1\!+\!\la_3\!)\!
\pm\!\sqrt{\!\left(\tan^2\!\!\beta (\la_2\!+\!\la_3\!)\!-\!
(\la_1\! +\!\la_3\!)\!\right)^2 \!+\!
4\la_3^2\tan^2\!\!\beta}\right].}

First, $c_-<0$ so that one of the Higgs fields approaches to the true vacuum
from the wrong direction. Thus $c_-$ is not a good solution. Now we are left
with $c_+>0$.

Second, if $\la_3\to 0$, $\tc$ either increases indefinitely or approaches
to $0$, unless the coefficient of the linear term in eq.\enorii\
vanishes. (This vanishing condition is nothing but eq.\eresult\
as $\la_3\to 0$.) This is unreasonable because
for vortex solutions with the same characteristic length
we should expect $\tc_1$ and $\tc_2$ are of a similar order
for any $\la_3$. Without loss of generality, suppose $c_1\ll c_2\sim 1$ for
$0<\la_3\ll 1$, then one can easily see that
the characteristic length $\xi_1$ fails to make sense for $\phi_1$.
Thus it is not a solution.
We believe that as an exact solution this will fail to satisfy the boundary
conditions at $r=0$.

Therefore, eq.\enorii\ cannot be solved for plausible $\tc$ consistently for
$\tan\beta$ as an input parameter.
 This leads us to the conclusion
 that we should treat $\tc$ as an input parameter.
Thus $\tc=1$, which leads to eq.\eresult\ consistently, is a
well-defined reasonable assumption to present our argument.

In the case of $\la_5\neq 0$, if I quote the result,
we also obtain an inequality as well
\eqn\edfvr{(\la_5 -\la_1)(\la_5-\la_2)>0.}
Following a similar argument, we obtain
\eqn\edresult{\tan\beta\equiv {v_2\over v_1}=\sqrt{{\la_1-\la_5\over
\la_2-\la_5}},\ \ \  \la_3\neq 0\ \ {\rm or} \ \la_5\neq 0.}
Thus we have determined the ratio of the two Higgs \VEV s in terms of the
couplings in the Higgs potential. This tells us that although different Higgs
field gets different \VEV s, their characteristic lengths should be the same to
form a single defect. Both Higgs should reach the true vacuum at the same
distance. To do that the two \VEV s should satisfy a proper relation, which is
Eq.\edresult.

Furthermore, together with $v$, we can completely determine the \VEV s as
\eqn\edIII{v_1 ={v\over\sqrt{2}}
\,\cos\beta={v\over\sqrt{2}}{\sqrt{{\la_2-\la_5\over \la_1+\la_2-2\la_5}}},\ \
v_2 ={v\over\sqrt{2}}
\,\sin\beta={v\over\sqrt{2}}{\sqrt{{\la_1-\la_5\over \la_1+\la_2-2\la_5}}}.}
The characteristic lengths $\xi_1, \xi_2$, now satisfy
\eqn\edchar{\xi\equiv\xi_1=\xi_2={1\over v}
\sqrt{{\la_1+\la_2-2\la_5\over \la_1\la_2+\la_2\la_3+\la_3\la_1
-\la_5(2\la_3+\la_5)}}.}
Note that, although $\tan\beta$ does not depend on $\la_3$, it is crucial to
have nonvanishing $\la_3$ or $\la_5$ coupling to obtain such a result.
The gauge boson mass is $M_Z=1/\la=\gt v$
after spontaneous symmetry breaking.

In this two-Higgs-doublet model there are five physical Higgs bosons:
$H^{\pm},\ A^0,$ $ \ H^0,\ h^0$. $A^0$ is a CP-odd neutral scalar, while
$H^0,\ h^0$ are CP-even scalars. $h^0$ denotes the lightest Higgs.
Using Eq.\edIII, we can compute the masses of all these physical Higgs
bosons in terms of the couplings in the Higgs potential and $v$, where
$v=247$ GeV. $\la_5$ is related to $M_{A^0}$ and $M_{H_0,h_0}$ can be
determined
in terms of $\la_1,\la_2,\la_3,\la_5$ and $v$. Thus we only have five free
parameters, if such an electroweak Z-string exists.

In particular the neutral Higgs masses become impressively simple:
\eqn\edhms{\eqalign{
m_{H^0} &=v^2\left[\la_3 + {\la_1\la_2-\la_5^2\over \la_1+\la_2
-2\la_5}\right],\cr
m_{h^0} &=v^2 {2\la_1\la_2-\la_5(\la_1+\la_2)\over 2(\la_1+\la_2
-2\la_5)}.\cr}}
Note that $m_{h^0}$ does not depend on $\la_3$.

The appearance of integers in the solutions,
which we still call ``winding'' number,
is rather intriguing because there is no
explicit $U(1)$ symmetry to be broken
which should determine the necessary topological sector.
If our vortex solutions are nontopological as in ref.%
\ref\tdlee{T.~D.~Lee, Phys. Rep. {\bf 23} (1976) 254.},
there should not be such a parameter.
This however
can be explained as follows: If we regard $W_\mu^1=0=W_\mu^2$ as gauge fixing
conditions, then effectively we can view
the symmetry of the system as $U(1)\times U(1)_Y$. When we twist this symmetry
to obtain $U(1)_{em}$, the remaining twisted $U(1)_{\gt}$
is spontaneously broken to lead to the winding sector.

Since $SU(2)$ is a simple group, $U(1)\times U(1)_Y$ is not an invariant
subgroup of $SU(2)\times U(1)_Y$.
Furthermore, $\pi_1\left(SU(2)\times U(1)_Y/U(1)_{em}\right) =0$
implies that these winding sectors would not provide any topological stability.
In other words, they must be gauge equivalent to $n=1$ solution via deformable
gauge transformation.

Even for $n=1$ solution it is most likely that this solution would not saturate
the Bogomol'nyi bound.
Although it obviously is a finite energy solution, it
does not seem to be a classically stable solution. It is argued that there is a
case of quantum stabilization of a classically unstable solution%
\ref\preskill{J. Preskill,  Caltech preprint, CALT-68-1787 (hep-ph/9206216).}.
It however cannot be applied in this case unless one of $\la_4$ or $\la_5$
vanishes because there is no tree level Goldstone boson.
If $\la_5=0$, then there is one Goldstone boson. But the argument still does
not apply because $\pi_1(G_{{\rm gauge}}/H_{{\rm gauge}})$ is still trivial.

\newsec{Discussion}

We have shown that two-Higgs systems in general admit vortex solutions, which
requires a specific ratio of the two VEVs. This can be understood more easily
if we use the condition
$$\widehat\phi\equiv{\phi_1\over v_1}={\phi_2\over v_2}.$$
Note that this condition not only shows up as a consistency condition of the
vortex solutions but also shows up as that of the monopole solution considered
in \Mydhiggs.

Then as we can easily see, the field equations reduce to the field equations of
a single Higgs with the VEV $v$ and the Higgs potential
$V(\phi)=\la (|\phi|^2-v^2/2)^2$, where $\phi=v\widehat{\phi}/\sqrt{2}$
and
$$\la\equiv {\la_1\la_2+\la_2\la_3+\la_3\la_1-\la_5(2\la_3+\la_5)
\over \la_1+\la_2-2\la_5}.$$
All such results can be obtained if
$$\tan\beta\equiv {v_2\over v_1}=\sqrt{{\la_1-\la_5\over
\la_2-\la_5}},\ \ \  \la_3\neq 0.$$
These results also apply to the cases of $\la_5=0$.

In the supersymmetric models $\la_1=\la_2$ at tree level. With the RGE
improvement, which leads to $\la_1\neq\la_2$,
we can show that small $\tan\beta$ is preferred  for the existence
of the electroweak Z-string, if the supersymmetry breaking scale is
sufficiently different from the electroweak breaking scale.
Under the current experimental constraint this is not unreasonable. The proton
decay constraint also prefers small $\tan\beta$\ref\ArNa{P. Nath and R.
Arnowitt, CTP-TAMU-27/92.}.

For the unstable solutions we need further investigation to find out what kind
of cosmological trace they could leave. But one certain thing is that as soon
as we find out how nature selects out $\tan\beta$ (if nature prefers the
two-doublet model), we shall understand its mystery if $\tan\beta$ turns out to
meet our claim.
The ultimate proof of the existence of such vacuum defects should be determined
by experiments or by observations.

Much work is needed to find exact solutions to determine the exact value of
$c$.
It will be also important and interesting to find out if there are any other
reasons nature prefers a specific $\tan\beta$ theoretically. This will be still
a question to be answered after we determine experimentally. In this context,
we
hope this work can provide a clue to future investigation.

\bigbreak\bigskip\bigskip\centerline{{\bf Acknowledgements}}\nobreak

\par\vskip.3truein

The author would like to thank R. Arnowitt
for helpful discussions.
This work was supported in part by NSF grant PHY89-07887 and World Laboratory.


%
\par\vskip.3truein

\footatend\immediate\closeout\rfile\writestoppt
\baselineskip=14pt\centerline{{\bf References}}\bigskip{\frenchspacing%
\parindent=20pt\escapechar=` \input refs.tmp\vfill\eject}\nonfrenchspacing
\vfill\eject
\bye